\documentclass[aps,prb,reprint,superscriptaddress]{revtex4-1}
\usepackage{amsmath,amssymb}
\usepackage{multirow}
\usepackage{graphicx}
\usepackage{epstopdf}
\usepackage{tabularx, booktabs}
\usepackage[titletoc,toc,title]{appendix}
\usepackage[colorinlistoftodos]{todonotes}
\usepackage{pbox}
\usepackage{siunitx}
\newcolumntype{Y}{>{\centering\arraybackslash}X}
\usepackage[caption=false,position=b,singlelinecheck=off,font=normalsize,labelfont=bf,justification=justified]{subfig}
\usepackage{hyperref}
\usepackage{adjustbox} 
\usepackage[version=4]{mhchem}
\hypersetup{
    colorlinks=true,
    linkcolor=blue,
    filecolor=black,      
    urlcolor=blue,                                                                       
    citecolor=blue,
}
\graphicspath{ {Figures/} }

\begin{document}
\title{Site selective spin and orbital excitations in \ce{Fe_3O_4 }}
\author{H. Elnaggar}
\email[]{H.M.E.A.Elnaggar@uu.nl}
\affiliation{Debye Institute for Nanomaterials Science, Utrecht University, Universiteitsweg 99, 3584 CA Utrecht, The Netherlands.}
\author{R. Wang}
\affiliation{Debye Institute for Nanomaterials Science, Utrecht University, Universiteitsweg 99, 3584 CA Utrecht, The Netherlands.}
\author{S. Lafuerza}
\affiliation{European Synchrotron Radiation Facility, CS40220, F-38043 Grenoble Cedex 9, France.}
\author{E. Paris}
\affiliation{Photon Science Division, Paul Scherrer Institut, Forschungsstrasse 111, 5232 Villigen PSI, Switzerland.}
\author{A. C. Komarek} 
\affiliation{Max-Planck-Institute for Chemical Physics of Solids, N\"{o}thnitzer Str. 40, 01187, Dresden, Germany.}
\author{ H. Guo} 
 \affiliation{Max-Planck-Institute for Chemical Physics of Solids, N\"{o}thnitzer Str. 40, 01187, Dresden, Germany.}
\author{Y. Tseng}
\affiliation{Photon Science Division, Paul Scherrer Institut, Forschungsstrasse 111, 5232 Villigen PSI, Switzerland.}
\author{D. McNally}
\affiliation{Photon Science Division, Paul Scherrer Institut, Forschungsstrasse 111, 5232 Villigen PSI, Switzerland.}
\author{F. Frati}
\affiliation{Debye Institute for Nanomaterials Science, Utrecht University, Universiteitsweg 99, 3584 CA Utrecht, The Netherlands.}
\author{M. W. Haverkort}
\affiliation{Institut f\"{u}r Theoritiche Physik, Universit\"{a}t Heidelberg, Philosophenweg 19, 69120 Heidelberg, Germany. }
\author{M. Sikora}
\affiliation{Academic Centre for Materials and Nanotechnology, AGH University of Science and Technology, Mickiewicza 30, 30-059, Krakow, Poland.}
\author{T. Schmitt}
\affiliation{Photon Science Division, Paul Scherrer Institut, Forschungsstrasse 111, 5232 Villigen PSI, Switzerland.}
\author{F.M.F. de Groot}
\email[]{F.M.F.deGroot@uu.nl}
\affiliation{Debye Institute for Nanomaterials Science, Utrecht University, Universiteitsweg 99, 3584 CA Utrecht, The Netherlands.}

\date{\today}

\begin{abstract}
\ce{Fe3O4} is a mixed-valence strongly correlated transition metal oxide which displays the intriguing metal to insulator Verwey transition. Here we investigate the electronic and magnetic structure of \ce{Fe3O4} by a unique combination of high resolution Fe $2p3d$ resonant inelastic scattering magnetic circular (RIXS-MCD) and magnetic linear (RIXS-MLD) dichroism. We show that by coupling the site selectivity of RIXS with the magnetic selectivity imposed by the incident polarization handedness, we can unambiguously identify spin-flip excitations and quantify the exchange interaction of the different sublattices. Furthermore, our RIXS-MLD measurements show spin-orbital excitations that exhibit strong polarization and magnetic field dependence. Guided by theoretical simulations, we reveal that the angular dependence arises from a strong interplay between trigonal crystal-field, magnetic exchange and spin-orbit interaction at the nominal \ce{Fe^{2+}} sites. Our results highlight the capabilities of RIXS magnetic dichroism studies to investigate the ground state of complex systems where in-equivalent sites and bonds are simultaneously present. 
\end{abstract}
\maketitle
\section{Introduction} 
%
Despite decades of research on magnetite (\ce{Fe3O4}), the first magnetic material known to man, numerous aspects of its physics remain unsolved. In particular, the origin of the metal to insulator transition (the Verwey transition at $T_V$ $\sim$ \SI{125}{\kelvin}) which results in a decrease of the electrical conductivity by two orders of magnitude is vastly debated. This drastic switch of conductivity was found to occur in the \textit{picosecond} timescale \cite{Pontius2011,deJong2013}. The ground state of \ce{Fe3O4} consequently poses interesting options for designing fast electronics. 
\\ 
\indent Above $T_V$,  \ce{Fe3O4} has a cubic inverse spinel crystal structure containing two different iron sites. \ce{Fe^{3+}} ions reside in tetrahedral ($T_d$) coordinated interstices (referred to as A sites) while both \ce{Fe^{2+}} and \ce{Fe^{3+}} ions are in octahedral ($O_h$) coordinated interstices (referred to as B sites). Verwey proposed that at $T_V$ an order to disorder transition takes place where the low temperature ordering of \ce{Fe^{3+}} and \ce{Fe^{2+}} ions at the B sites melt permitting relatively easy valencey exchange by means of fast electron hopping \cite{Verwey1939}. Ever since this formulation, extensive efforts were exerted to find evidence of the proposed charge ordering (and later on, orbital ordering) and the accompanying distortion at the low temperature phase \cite{Hamilton1958,Yamada1968,Hargrove1970,Gasparov2000,Novak2000,Garcia2001,Gracia2004a,Senn2012b,Senn2013,Bosak2014,Shin2015,Huang2017}.  Unfortunately, no consensus has been made regarding the charge or orbital ordering in the low temperature phase \cite{Walz2002,Garcia2004}.
\\
\indent  A key aspect to understand the physics of the Verwey transition and the complex interplay of various degrees of freedom in the low temperature phase is the determination of the ground state of the high temperature phase. Resonant inelastic X-ray scattering (RIXS) at the L-edges of transition metals ($2p \rightarrow 3d$ excitations) is a powerful tool sensitive to charge, orbital, spin and lattice degrees of freedom \cite{Ament2011,Bisogni2016a,Lu2018}. The chemical selectivity provided on resonance enables RIXS to probe independently the electronic properties stemming from different sites in a mixed-valence system. RIXS measurements performed using a specific incident X-ray polarization strongly restrict the accessible excited and final states providing additional discrimination power. The difference between the RIXS cross-section measured with circular left and right polarized incident X-rays on a magnetic sample (RIXS-MCD) provides information about the magnetic properties of the absorbing species \cite{Miyawaki2017}. A detailed study of the RIXS cross-section as a function of the angle of the incident linearly polarized X-rays (RIXS-MLD) probes sensitively the interplay between local distortion, spin-orbit and exchange interaction. 
\\
\indent In this work, we employed a combination of high resolution Fe $2p3d$ RIXS-MCD and RIXS-MLD in a \ce{Fe3O4} single crystal to probe low energy spin and spin-orbital excitations associated with the different Fe sites. We identified spin-flip excitations at the A and B sublattices by harnessing the magnetic contrast offered by RIXS-MCD and quantified the effective exchange interaction respectively. We determined the local site distortion at the \ce{Fe^{2+}} sites by a detailed study of the the RIXS-MLD angular dependence of the spin-orbital excitations. Through these unique set of experiments, we construct a comprehensive picture regarding the ground state of Fe in \ce{Fe3O4} above the Verwey transition which has been long disputed.

\section{Experimental details}
\begin{figure}[!htb]
    \begin{center}
    \adjustimage{max size={1\linewidth}}{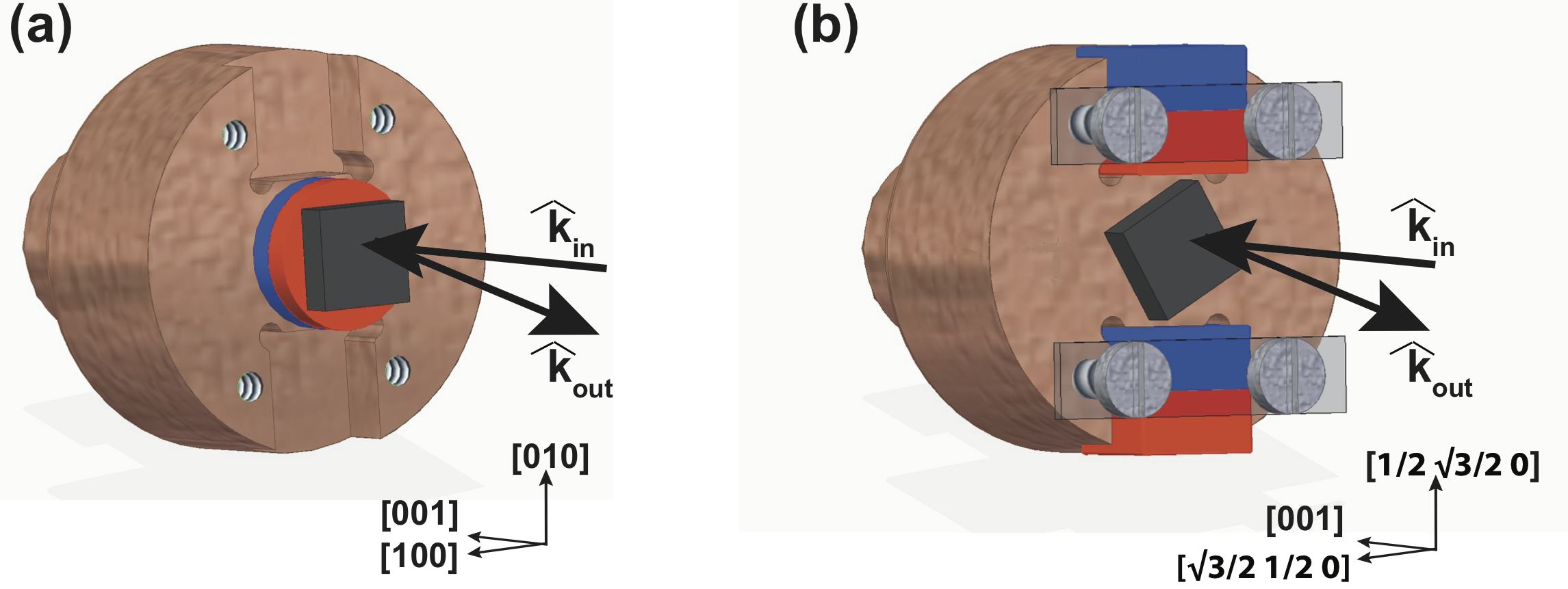}
    \end{center}
	\caption{Experimental setup. (a) Magnetic circular dichroism configuration (MCD). Here the external magnetic field is aligned parallel to the incident wave vector ($\boldsymbol{\widehat{k}_{in}}$) which corresponds to the $[001]$ crystallographic direction. (b) Magnetic linear dichroism configuration (MLD) where the external magnetic field is aligned perpendicular to $\boldsymbol{\widehat{k}_{in}}$ which corresponds to the $[1/2,\sqrt{3}/2,0]$ crystallographic direction. The axes system in the bottom shows the orientation of the crystal for both configurations.} 
	\label{Fig:Experimental_1}
\end{figure}

\indent We studied a high quality $(001)$ \ce{Fe3O4} single crystal grown at the Max-Planck Institute for chemical physics of solids. The stoichiometry of the single crystal was examined by performing temperature-dependent magnetic measurements. A sharp transition was observed at \SI{125}{\kelvin} as expected for a highly stochiometric sample \cite{supplementary}. The crystallinity and homogeneity of the sample was assessed by measuring the Darwin width of the $(008)$ reflection  across the sample at beamline ID28 of the European synchrotron radiation facility \cite{supplementary}. In order to clean and remove damaged surface layers, we exposed the surface to boiling concentrated \ce{HCl} for $\sim$ \SI{30}{\second} before the measurement.   
\\
\indent Fe L-edge X-ray absorption (XAS) and resonant inelastic X-ray scattering (RIXS) measurements were carried out at the ADRESS beamline of the Swiss Light Source at the Paul Scherrer Institut, Switzerland \cite{Strocov2010}. The high-brilliance X-ray beam was monochromatized using a plane grating with a constant groove density of \SI{800}{\mm^{-1}} and focused down to a spot of $\leq$ \SI{4x55}{\mu m} size at the sample position using an elliptical refocusing mirror. All measurements were performed at normal incidence, $i.e.$ with the incoming beam impinging at an angle of \ang{90;;} with respect to the sample surface. The scattering angle was set to $2 \theta =$ \ang{130;;} to minimize the elastic scattering angular dependence. The energy analysis of the emitted radiation was performed using a variable-line-spacing (VLS) spherical grating, with an average groove density of \SI{1500}{\mm^{-1}}, dispersing the emitted radiation onto a high-resolution CCD camera. The combined energy resolution was $\sim$ \SI{76.2}{\meV} full width half maximum (FWHM), determined by collecting the elastic scattering from a carbon tape reference \cite{supplementary}. The radiation source is a fixed-gap Apple-II type undulator \cite{Sasaki1993}, producing left and right circular polarized light as well as linear polarized light. The angular direction of the linear polarization can be varied continuously, from horizontal to vertical with respect to the scattering plane. The X-ray absorption spectra were recorded in total electron yield mode.
 \\
\indent A permanent gold coated \ce{NdFeB} magnet with a magnetic flux density on the surface of \SI{0.4}{\tesla} was used to saturate the magnetization \cite{supplementary}. The magnet was placed in two configurations for the measurements: (i) Magnetic circular dichroism configuration where the external magnetic field was aligned parallel to the incident wave vector ($\boldsymbol{\widehat{k}_{in}}$) (\textit{i.e.} in the scattering plane) which corresponds to the $[001]$ crystallographic direction (Fig. \ref{Fig:Experimental_1}a) and (ii) Magnetic linear dichroism configuration where the external magnetic field was aligned perpendicular to $\boldsymbol{\widehat{k}_{in}}$ (\textit{i.e.} out of the scattering plane) which corresponds to the $[1/2,\sqrt{3}/2,0]$ crystallographic direction (Fig. \ref{Fig:Experimental_1}b). This low symmetry direction was chosen to investigate magneto-crystalline anisotropy at the Fe sites of \ce{Fe3O4}.  All measurements were performed at fixed temperature of \SI{170}{\kelvin} (\textit{i.e.} above the Verwey transition in the cubic phase). To avoid any radiation damage effects, we moved to a fresh spot on the sample every \SI{10}{\minute}. During this time period no signs of radiation damage were detected \cite{supplementary}. 
\\
\section{Calculation details}
\indent We used the quantum many-body program Quanty \cite{Haverkort2012,Lu2014,Haverkort2014} to simulate Fe $2p3d$ RIXS magnetic circular and linear dichroism in \ce{Fe3O4}. Three independent cluster calculations were performed to account for the nominal \ce{Fe^{2+}} in nearly octahedral symmetry (\textit{\ce{O_h}}),  \ce{Fe^{3+}} in \textit{\ce{O_h}} symmetry and \ce{Fe^{3+}} in tetrahedral symmetry (\textit{\ce{T_d}}). The Hamiltonian used for the calculations consists of the following terms: (i) Coulomb interaction, (ii) crystal field potential, (iii) spin-orbit coupling and (iv) magnetic exchange interaction. The $d-d$ ($p-d$) multipole part of the Coulomb interaction was scaled to 70\%  (80 \%) of the Hartree-Fock values of the Slater integral. 
\\
\indent We investigated two possible distortion scenarios at the \ce{Fe^{2+}} sites, namely: tetragonal ($D_{4h}$) and trigonal ($D_{3d}$) local distortions. Tetragonal distrotion arising from trimeron correlations was simulated by summing up the contributions of three tetragonally distorted \ce{Fe^{2+}} clusters along the cubic $\boldsymbol{ \widehat{a}}$, $\boldsymbol{ \widehat{b}}$ and $ \boldsymbol{\widehat{c}}$ axes weighted by the ratio 5:5:6 as reported by X-ray diffraction refinement results \cite{Senn2012b}. We simulated trigonal distortion at the \ce{Fe^{2+}} sites by summing up the contributions of four trigonally distorted clusters along the $[111]$, $[\overline{1}11]$, $[1\overline{1}1]$ and $[11\overline{1}]$ cubic directions \cite{Grave1993}. This preserves the global cubic symmetry of \ce{Fe3O4}. The main parameters used for the calculations are reported in Table \ref{tab:CalculationParameters} and further information is provided in the Supplementary Materials \cite{supplementary}. The general parameters used for the simulations are in agreement with previous studies of \ce{Fe3O4} $L_{2,3}$ edges \cite{Kuiper1997,Pattrick2002,Arenholz2006a,Carvallo2008,Liu2017}. 
\\
\indent We aligned the incident energy of the calculation by fitting the experimental XMCD signal \cite{supplementary}. The relative shift between \ce{Fe^{3+}} and \ce{Fe^{2+}} B-sites was found to be \SI{0.7}{\eV} in agreement with previous elaborate investigations \cite{Liu2017}. This value is smaller than the expected chemical shift between \ce{Fe^{3+}} and \ce{Fe^{2+}} ions in a pure-valence system. This observation is also confirmed by X-ray diffraction measurements where the average bond lengths for both the \ce{Fe^{3+}} and \ce{Fe^{2+}} ions were found to be very similar \cite{Senn2012b}. 
\begin{center}
\begin{table}[h!]
\begin{tabular}{|c||c|c|c|}
\hline 
Parameter (eV)& \ce{Fe^{2+}_B}  & \ce{Fe^{3+}_B} & \ce{Fe^{3+}_A} \\
\hline  \hline
$10D_q$ ($O_h$)  & 1.20 & 1.20 &  -0.77\\ 
$D_s$  ($D_{4h}$) & -0.067 & 0 &  0 \\ 
$D_\sigma$ ($D_{3d}$) & 0.067 & 0 &  0 \\ 
Mean exchange field  ($J^*$)& 0.09& 0.09& -0.09\\ 
Slater integrals ($F^{2}_{dd}$,$F^{4}_{dd}$) & 7.68,4.77 & 7.67, 5.13 &7.67, 5.13 \\
\hline
\end{tabular} 
\caption{Main parameters used for the Fe $2p3d$ RIXS cluster calculations. The distortion parameters $D_s$ and $D_{\sigma}$ were applied to investigate the effect of tetragonal and triginal distortion at the \ce{Fe^{2+}} sites respectively. The subscripts used for the Fe ions refer to the A and B sites.} \label{tab:CalculationParameters}
\end{table}
\end{center} 
%
%
%
\section{Site selective spin-flip excitations}
\begin{figure}[!htb]
    \begin{center}
    \adjustimage{max size={1\linewidth}}{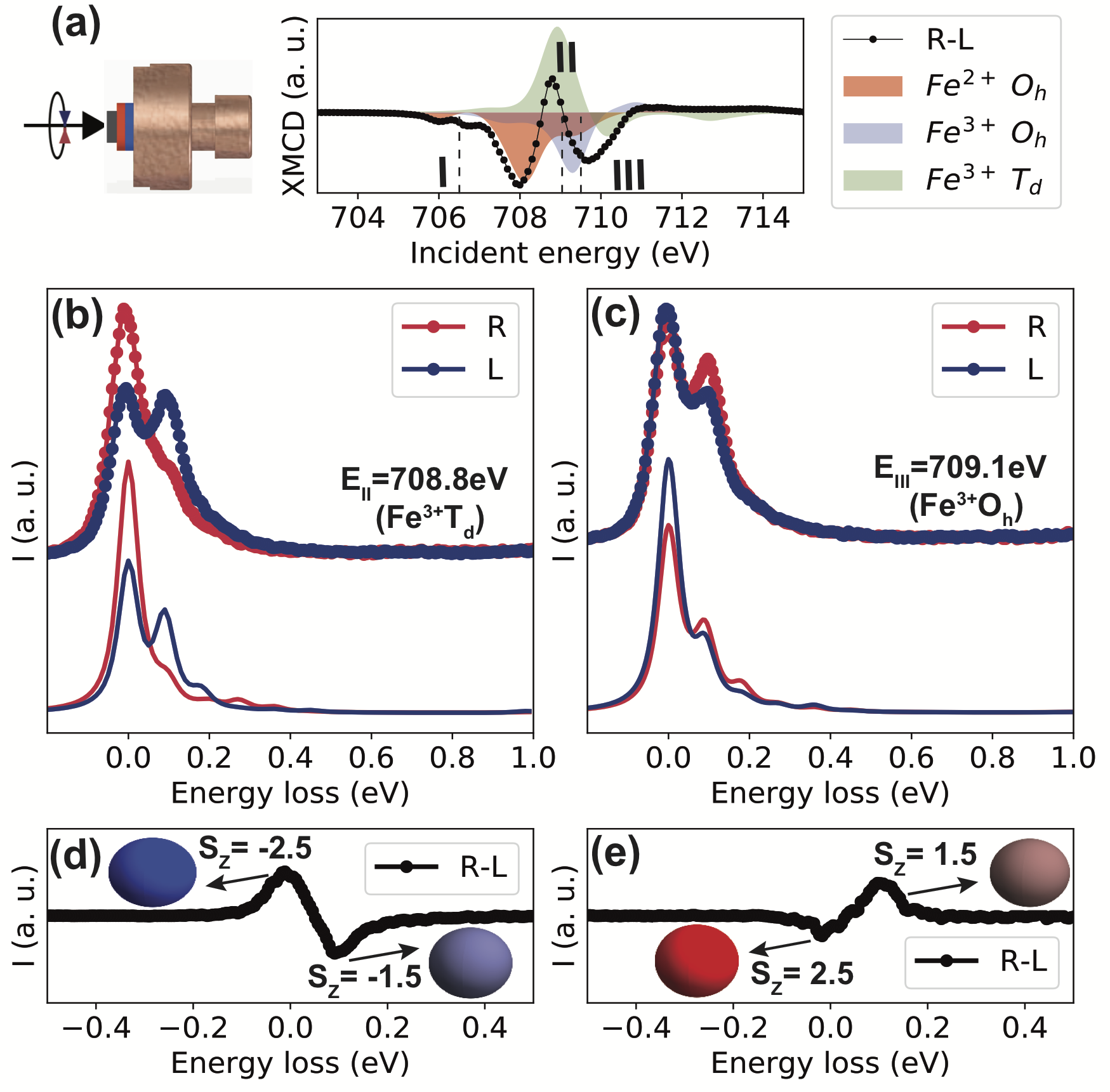}
    \end{center}
	\caption{ (a) Fe $L_3$ XMCD signal in \ce{Fe3O4}. The contributions of the three Fe ions to the XMCD signal are shown in filled colors. The experimental Fe $2p3d$ RIXS  cuts at $\boldsymbol{E_{II}}=$ \SI{708.8}{\eV} and at  $\boldsymbol{E_{III}}=$ \SI{709.1}{\eV} measured with right (R) and left (L) polarized X-rays are presented in panels (b) and (c) respectively. The \ce{Fe3O4} RIXS simulations at the two incident energies are shown vertically displaced with respect to the experiment in solid lines. (d) RIXS-MCD signal at $\boldsymbol{E_{II}}$. (e) RIXS-MCD signal at $\boldsymbol{E_{III}}$ for the \ce{Fe_B} sublattice. The charge density plots for the ground state and the first excited state are presented in the figures d and e. The intensity of the blue (red) color is proportional to the spin down (up) expectation value.} 
	\label{Fig:SpinFlip_1}
\end{figure}

\begin{figure}[h!]
    \begin{center}
    \adjustimage{max size={1\linewidth}}{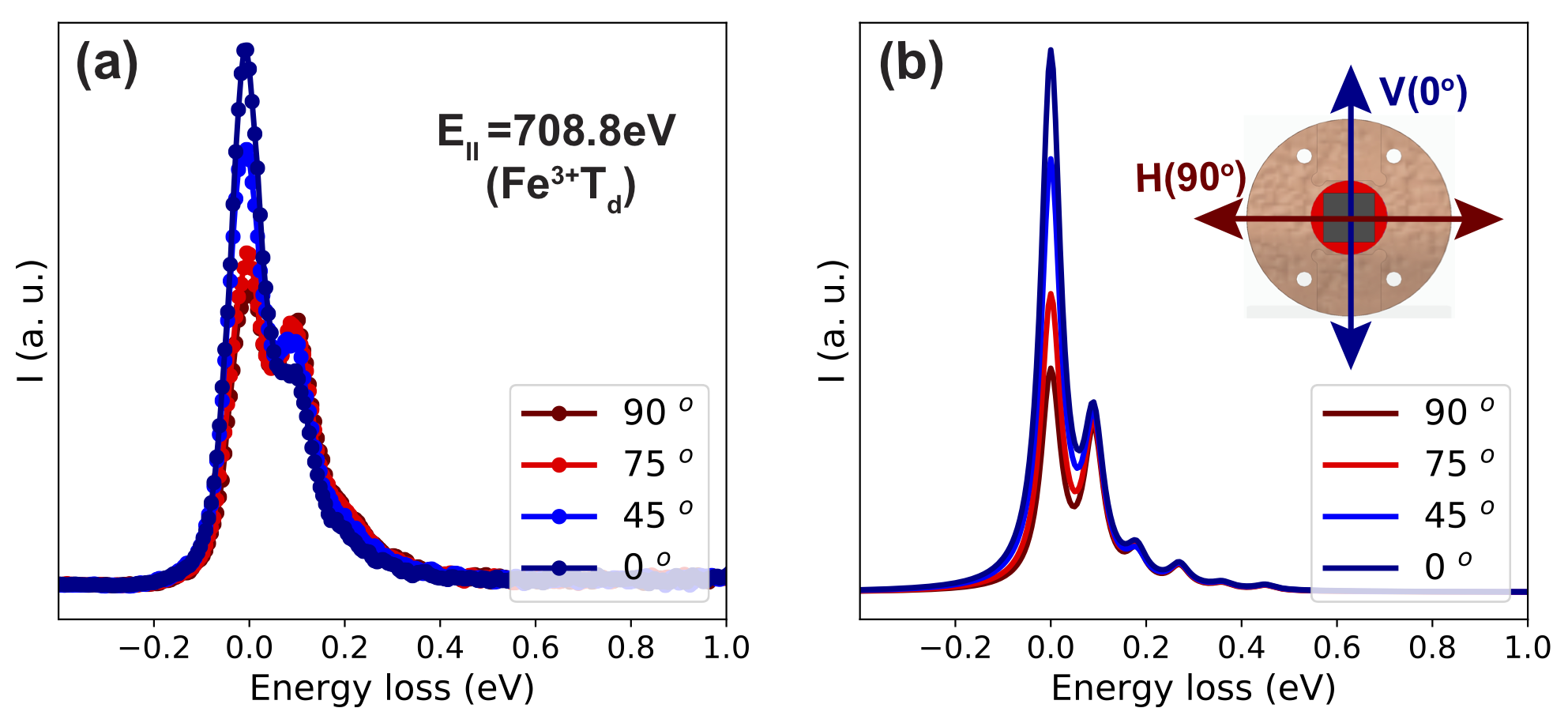}
    \end{center}
	\caption{Fe $2p3d$ RIXS magnetic linear angular dependent results at $\boldsymbol{E_{II}}=$\SI{708.8}{\eV} measured with the external magnetic field parallel to the incident wave vector ($\boldsymbol{\hat{k}_{in}}$). (a) Experimental RIXS cuts measured as a function of the rotation angle of the incident polarization vector ($\boldsymbol{\widehat{\epsilon}_{in}}$). The angle is defined from the vertical direction labeled \textbf{V}.  (b) Calculations of the Fe $2p3d$ RIXS magnetic linear angular dependence at $\boldsymbol{E_{II}}$ in \ce{Fe3O4}.  A sketch of the setup in the configuration referred to as MCD in the text is shown in the upper right corner.}
 \label{Fig:SpinFlip_3}
\end{figure}

\begin{figure}[h!]
    \begin{center}
    \adjustimage{max size={1\linewidth}}{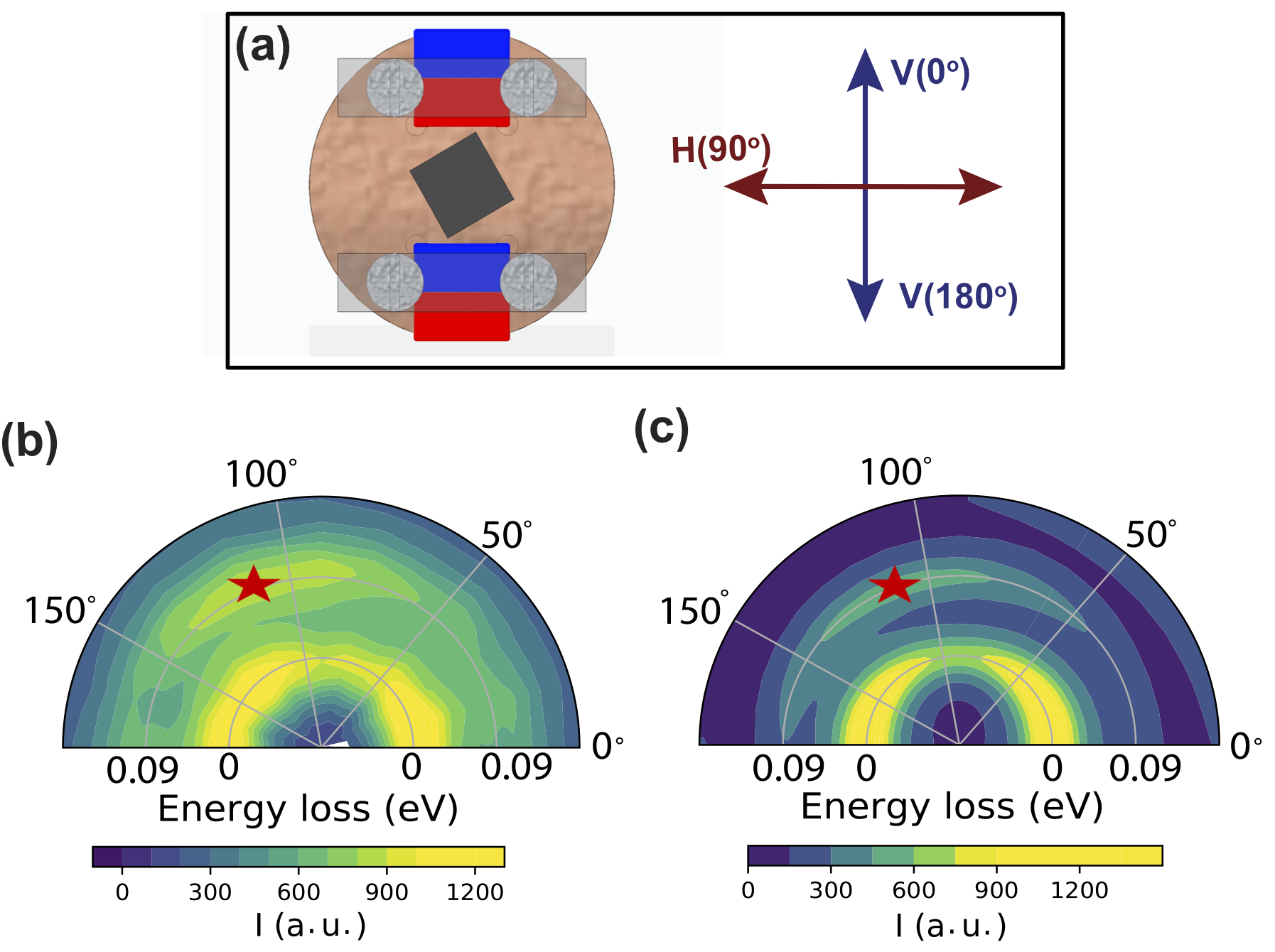}
    \end{center}
	\caption{Fe $2p3d$ RIXS magnetic linear angular dependent results at $\boldsymbol{E_{II}}=$\SI{708.8}{\eV} measured in the MLD configuration. (a) Sketch of the setup in the configuration referred to as MLD in the text. (b) Experimental RIXS data measured as a function of the rotation angle of the incident polarization vector ($\boldsymbol{\widehat{\epsilon}_{in}}$). The angle is defined from the vertical direction labeled \textbf{V}. The data is presented in a polar contour plot where the radial axis gives the energy loss and the polar angle gives the angle between $\boldsymbol{\widehat{\epsilon}_{in}}$ and the vertical direction. (c) Calculations of the Fe $2p3d$ RIXS magnetic linear angular dependence at $\boldsymbol{E_{II}}$ in \ce{Fe3O4}. The maximum intensity of the spin-flip feature is marked with a red star for a visual guide.}
 \label{Fig:SpinFlip_2}
\end{figure}

\indent To identify the magnetic spectral features of the three Fe ions in \ce{Fe3O4}, Fe \ce{L_{2,3}} XMCD measurements were performed (Fig. \ref{Fig:SpinFlip_1}a). The results agree well with previous measurements \cite{Pattrick2002,Goering2005,Arenholz2006a,Goering2007,Antonov2003,Carvallo2008} and confirm the good stoichiometry and surface quality of the sample \cite{supplementary}. We identify three main XMCD features that we will focus on further in this paper: \textbf{I} at \SI{706.1}{\eV}, \textbf{II} at \SI{708.8}{\eV} and \textbf{III} at \SI{709.1}{\eV} (see Fig. \ref{Fig:SpinFlip_1}a). Based on the XMCD calculations, tuning the incident energy to $\boldsymbol{E_{II}}$ maximizes the magnetic signal of the \ce{Fe^{3+}} $T_d$ sites, while at  $\boldsymbol{E_{III}}$ the signal of \ce{Fe^{3+}} $O_h$ sites is maximized.
\\
\indent RIXS measurements at $\boldsymbol{E_{II}}$ ($\boldsymbol{E_{III}}$) are presented in Fig. \ref{Fig:SpinFlip_1}b (c). The lowest energy loss feature observed for both incident energies is centered at \SI{90}{\meV}. A strong RIXS-MCD signal is observed for both cuts emphasizing the magnetic nature of the feature. Most importantly, the RIXS-MCD signal reverses sign from $\boldsymbol{E_{II}}$ to $\boldsymbol{E_{III}}$. The origin of the sign reversal can be understood by a close inspection of the RIXS process. The RIXS cross-section can be expressed as a sum of fundamental spectral functions multiplied with a geometry tensor involving the polarization of the incident and scattered photons \cite{Juhin2014}. The fundamental spectra can be grouped according to their symmetry into an isotropic contribution ($\sigma ^{(0)}$), an MCD active contribution ($\sigma ^{(1)}$) and an MLD active contribution ($\sigma ^{(2)}$) \cite{Haverkort2010}. Consequently, the $2p3d$ RIXS-MCD cross-section is controlled by the XMCD spectral function and the sign reversal stems from the incident energy site selectivity. Hence, by measuring Fe $2p3d$ RIXS at incident energies \textbf{II} and \textbf{III} we can probe selectively the exchange interaction of the \ce{Fe_A} and \ce{Fe_B} sublattices in \ce{Fe3O4}.
\\
\indent Theoretical calculations based on the single site impurity model with an effective exchange field $J^{*}=$\SI{90}{\meV} reproduces the experimental data well (Fig. \ref{Fig:SpinFlip_1}b and c). The \SI{90}{\meV} peak can be assigned to a spin-flip excitation enabled by the core-hole spin-orbit coupling in the intermediate state \cite{DeGroot1998,Haverkort2010}. At $\boldsymbol{E_{II}}$, a single spin-flip excitation ($\langle \Delta S_z \rangle = 1 \hbar$) at tetrahedrally coordinated \ce{Fe^{3+}} sites occurs (see Fig. \ref{Fig:SpinFlip_1}d) while at $\boldsymbol{E_{III}}$ it corresponds to a single spin-flip excitation ($\langle \Delta S_z \rangle= -1 \hbar $) at octahedrally coordinated \ce{Fe^{3+}} sites (see Fig. \ref{Fig:SpinFlip_1}e). Our measurements show a shoulder corresponding to a double spin-flip excitation ($\langle \Delta S_z \rangle = 2 \hbar$ as shown in the Supplementary \cite{supplementary}) however we do not observe triple or higher excitations as the intensity of these peaks becomes significantly low and they are expected to be very broad band like excitations due to multiple possibilities to propagate the transferred momentum \cite{Haverkort2010,Betto2017}. 
\\
\indent To gain insight into the second rank tensor ($\sigma ^{(2)}$) holding information about site anisotropy, we measured $2p3d$ RIXS-MLD. The RIXS magnetic linear angular dependence was measured by rotating the incident polarization ($\boldsymbol{\widehat{\epsilon}_{in}}$) with respect to the crystallographic direction and the external magnetic field. We initially measured the RIXS-MLD signal with the magnetic field parallel to $\boldsymbol{\hat{k}_{in}}$ (\textit{i.e.} in the MCD configuration) as a reference measurement. In this case, the angle between $\boldsymbol{\widehat{\epsilon}_{in}}$ and the magnetic moments does not change as a function of rotation and hence almost no angular dependence is expected for the spin-flip excitation as confirmed by our measurements and calculations (Fig. \ref{Fig:SpinFlip_3}). Only the elastic peak exhibits angular dependence due to the zero's order elastic scattering cross-section. In contrast, a strong angular dependence is observed when we place the external magnetic field out of the scattering plane (the magnetic linear dichroism configuration) as shown in Fig. \ref{Fig:SpinFlip_2}. Here the angle between $\boldsymbol{\widehat{\epsilon}_{in}}$ and the magnetic moment changes as a function of rotation leading to a strong angular dependence. 
\\
\indent We investigated the coupling of the spin and orbital degrees of freedom at the \ce{Fe} sites of \ce{Fe3O4} by displacing the external magnetic field \ang{30;;} from the high symmetry [010] direction. Orienting the external magnetic field in a low symmetry direction aligns the net spin moment parallel to field.  If the orbital moments are not fully quenched, they consequently also re-align towards the low symmetry direction. The final orientation of the net magnetic moment depends on the strength of the competing interactions such as spin-orbit coupling and distortion. Hence the phase shift of the maximum intensity of the spin-flip excitation can be used to quantify magnetic-moment-induced distortion of the electron cloud. Based on this concept, we explored the possibility of the presence of orbital moments at the \ce{Fe_A} sites. Nominal tetrahedrally coordinated \ce{Fe^{3+}} sites have no orbital moment ($3d^5$ ion), however it has been reported that a strong non-vanishing orbital moment is present at the \ce{Fe_A} sites due to major charge transfer from the neighbouring oxygen \cite{Goering2011}. Our measurements show that the maximum of the RIXS magnetic angular dependence is at $\sim$ \ang{120;;}. This angle corresponds purely to the sample displacement (\ang{90;;}+\ang{30;;}) and no evidence of orbital moment induced shifts are found. The angular dependence follows that expected for $\boldsymbol{l_z} \approx 0$ case as can be seen by comparing the experiment to  the calculations in Fig. \ref{Fig:SpinFlip_2}c.
\\
\indent We remark that our spin-flip assignment is also consistent with inelastic neutron scattering measurements which show two spin wave modes ($ \Delta_5 $ and $ \Delta_2^{'}$) at $\sim$ \SI{80}{\meV} that nearly do not disperse \cite{McQueeney2007}. These modes were predicted to propagate solely on the \ce{Fe_B} sublattice in agreement with our site selectivity at $\boldsymbol{E_{III}}$. Another mode was observed at $\sim$ \SI{125}{\meV} which propagates on the \ce{Fe_A} sublattice and also has a very weak dispersion. We expect to probe this mode by tuning our incident energy to \textbf{II}, however our current experimental resolution is not enough to establish the displacement of the peak with \SI{35}{\meV}. Nevertheless, the reversal of the RIXS-MCD signal is a strong proof of probing the \ce{Fe_A} sublattice and confirms the site and magnetic selectivity of our experiment. The weak dispersion of the two modes is too small to be identified with current state of the art $2p3d$ RIXS instrumental resolution but it is foreseen that with future developments this could be probed like the progress seen in the case of nickelates \cite{DeGroot1998,Ghiringhelli2009,Betto2017,Lu2018}. Although phononic excitations were also reported at $\sim$ \SI{83}{\meV} \cite{Gasparov2000}, the strong magnetic circular and linear dichrosim observed together with the excellent agreement of the experimental data with simulations makes it an unlikely assignment. In addition, O K edge RIXS measurements showed \SI{70}{\meV} excitation that is related to phonons suggesting that the \SI{90}{\meV} is mainly magnetic in origin \cite{Huang2017}.
\section{Site selective spin-orbital excitations}
\begin{figure}[h!]
    \begin{center}
    \adjustimage{max size={1\linewidth}}{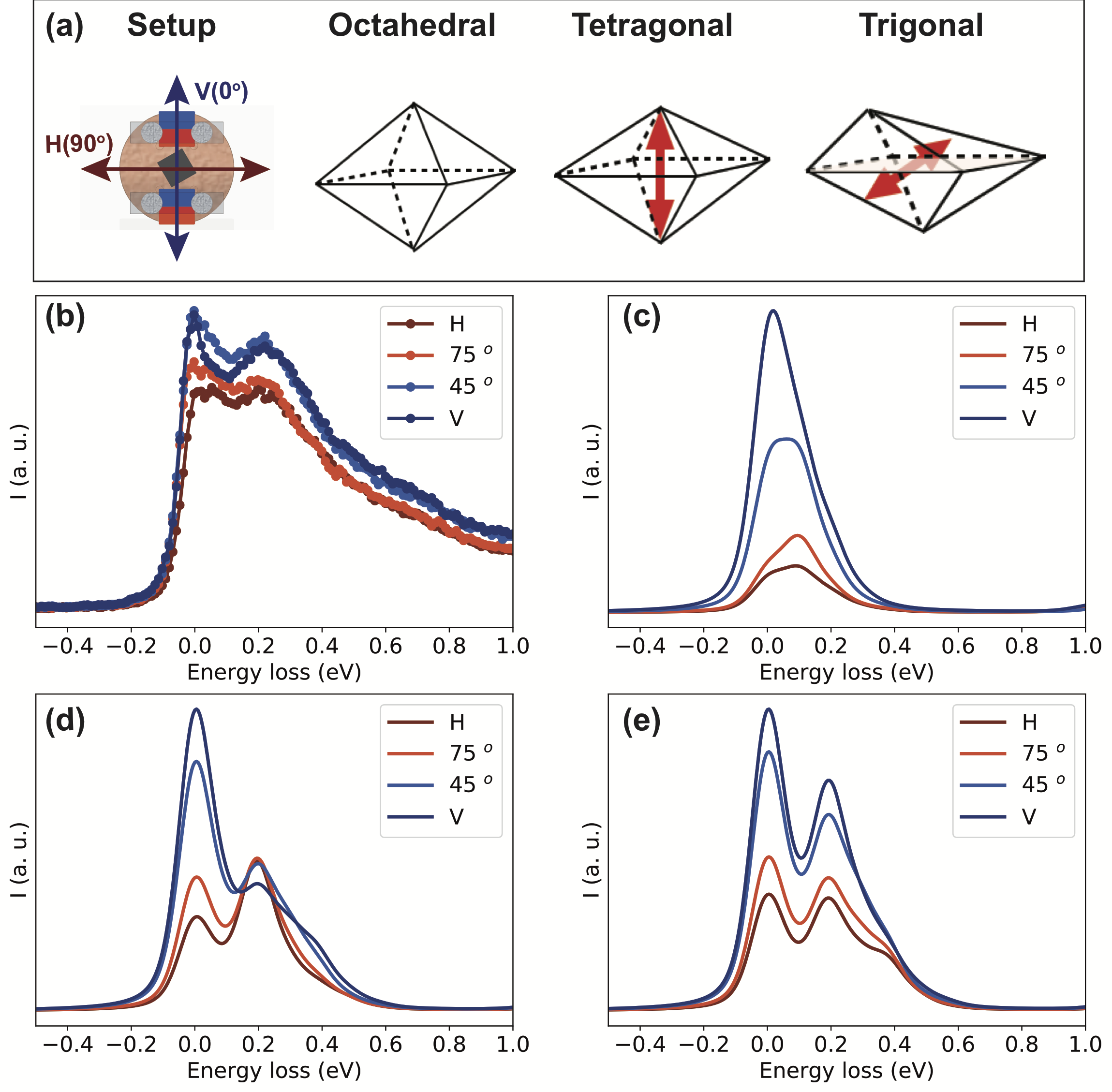}
    \end{center}
	\caption{Fe $2p3d$ RIXS magnetic linear angular dependent  results at  $\boldsymbol{E_{I}}$=\SI{706.1}{\eV}. (a) Sketch of the experimental MLD-configuration setup and the three local sites symmetries for the simulations presented in c, d and e respectively. (b) Experimental RIXS angular dependence measured as a function of the angle of the incident polarizaton vector ($\boldsymbol{\widehat{\epsilon}_{in}}$) in the MLD configuration. The angle is defined from the vertical direction labeled \textbf{V}. Calculations of the Fe $2p3d$ RIXS angular dependence in \ce{Fe3O4} with nominal \ce{Fe^{2+}} in: (c) $O_h$ symmetry, (d) tetragonal symmetry according to that reported by Senn \textit{et. al.} \cite{Senn2012b} and (e) in trigonal symmetry.} \label{Fig:Distortion_1}
\end{figure}

\indent In the following section we focus on the RIXS-MLD signal at $\boldsymbol{E_{I}}$ where we are sensitive to the nominal \ce{Fe^{2+}} sites. A broad low energy loss peak is observed at \SI{200}{\meV} (Fig. \ref{Fig:Distortion_1}b) which cannot be interpreted as a pure magnetic excitation (a double-spin excitation propagating on the \ce{Fe_B} sublattice) as illustrated by our calculations. The low energy feature can be best interpreted as a $dd$ excitation. In a pure octahedral coordination, the weak spin-orbit (\SI{66.5}{\meV}) and exchange (\SI{90}{\meV}) interaction cannot split the 15 fold $^5T_{2g}$ ground state enough to produce a pronounced peak centered at \SI{200}{\meV} (Fig. \ref{Fig:Distortion_1}c). A symmetry reduction of the \ce{Fe^{2+}} sites is necessary to reproduce the experimental spectra. We emphasize that such weak distortions does not significantly modify the XAS and XMCD spectral shape and hence can only be effectively studied with RIXS measurements (The effect of distortion on the XMCD calculations is presented in the Supplementary Materials \cite{supplementary}).  
\\
\indent A plausible scenario for such a symmetry lowering at the \ce{Fe^{2+}} sites is the presence of fluctuating trimeron correlations inherited from the low temperature phase that induces dynamical local tetragonal distortion above $T_V$. This tetragonal compressive distortion is reported to be along the $<100>$ axes of the high temperature cubic phase in a 5:5:6 ratio giving rise to a small anisotropy along the $\boldsymbol{ \widehat{c}}$ axis \cite{Senn2012b}. Fig. \ref{Fig:Distortion_1}d shows the calculated RIXS angular dependence for this trimeron-type distortion. We note that the horizontal polarization result agrees well with the experiment, however, the full angular evolution is inconsistent with the trimeron scenario. This partial agreement explains the assignment of the \SI{200}{\meV} feature to tetragonal distortion in previous work based on the results of horizontal polarization measurements \cite{Huang2017} and highlights the importance of detailed magnetic angular dependence measurements. 
\\
\indent Another candidate for the symmetry lowering is the trigonal distortion at the \ce{Fe^{2+}} sites along the $<111>$ axes of the high temperature cubic phase. Simulations using trigonal distortion capture the angular evolution of the RIXS cross-section well as illustrated in Fig. \ref{Fig:Distortion_1}e. A minor discrepancy between the calculated and experimental elastic line intensity can be seen which we attribute to self-absorption. Self-absorption of the emitted X-rays dominates at zero energy loss in particular when the incident energy is tuned at the pre-edge. This leads to the decrease of the elastic intensity with respect to the simulations. The details of the ground state determination are presented in the Supplementary \cite{supplementary}. Our current interpretation of the distortion is in agreement with results reported from M\"{o}ssbauer spectroscopy \cite{Grave1993} and diffraction anomalous fine structure \cite{Garcia2000}.
\\
\indent In view of the diverging interpretations of the Verwey transition \cite{Walz2002,Garcia2004}, we address hereafter the implication of our finding in relation to the electron transport mechanism in \ce{Fe3O4}. The type of distortion at the nominal \ce{Fe^{2+}} sites is a crucial piece of information that can establish a specific conduction mechanism. Evidence of dynamical local tetragonal distortion at the \ce{Fe^{2+}} sites in the high temperature phase would confirm, beyond any doubts, the high temperature polaron transport mechanism (\textit{i.e.} thermally activated hopping of localized electrons). Previous Fe $2p3d$ RIXS measurements reported that the \ce{Fe^{2+}} sites are tetragonally distorted and interpreted this as evidence of dynamical trimeron magnetic-polaronic correlations in the high temperature phase\cite{Huang2017}. Our refined experiment differs with this finding based on the detailed angular dependence and establishes that the feature at $\sim$ \SI{200}{\meV} is a result of trigonal distortion. This trigonal distortion is consistent with the point-group symmetry of the Fe B-sites and the magnitude of the distortion we find ($D_\sigma$ = \SI{67}{\meV}) is  in agreement with the static crystal-field splitting we find from DFT calculations. However, we do not completely dismiss the role of high temperature polaronic correlations as the distortion magnitude lies well within the phonon energies of \ce{Fe3O4} as discussed by Haupricht \textit{et. al.} in the case of \ce{Fe^{2+}} impurities in \ce{MgO} thin films \cite{Haupricht2010}. Our results proof that such correlations, if existing, do not induce tetragonal distortion along the cube axes but rather trigonal distortion along the cube diagonals. This type of dynamical correlations have been also supported by neutron diffuse scattering \cite{Yamada1979}, X-ray diffuse scattering \cite{Bosak2014} and some theoretical models \cite{Uzu2006,Uzu2008}. We expect that performing RIXS-MLD angular dependence at the low temperature phase on detwinned micro-crystals \cite{Senn2012b} or thin films \cite{Tanaka2012} in search for a symmetry change from trigonal to tetragonal distortion at the Verwey transition would provide deep insights in the complex physics of \ce{Fe3O4}.  \\
\\
\section{Conclusion}
\indent In conclusion, we studied low-energy spin and spin-orbital excitations associated with the different Fe sites in \ce{Fe3O4} using a combination of RIXS-MCD and RIXS-MLD measurements at the Fe $L_3$ edge. The magnetic discrimination emphasized by MCD offers a powerful way to distinguish magnetic excitations from other low-energy excitations (such as vibrational and orbital excitations). This technique was exploited on \ce{Fe3O4} to disentangle spin-flip excitations at the \ce{Fe_A} and \ce{Fe_B} sublattices. Our study demonstrates the capabilities of RIXS magnetic dichroism to determine site-selectively the exchange interaction in strongly correlated systems such as multiferroic systems, frustrated magnets and superconductors where identifying magnetic and lattice excitations is a crucial aspect. In particular, our methodology offers a possibility to thoroughly investigate spin excitations in thin films and multilayers that cannot be studied with conventional techniques as neutron scattering. In contrary to previous reports, we showed through detailed angular dependent RIXS measurements that the local symmetry of the nominal \ce{Fe^{2+}} sites is trigonally distorted. This finding proves that more complex correlations than the reported low temperature trimerons can be present in the high temperature phase.
\section{Acknowledgment}
\indent We thank A. Bosak and M. Hussein for their help with the X-ray diffraction measurements and sample characterization. A. van der Eerden and S. Deelen are thanked for their help with designing the sample holder for the synchrotron measurements. We are grateful for the fruitful discussions with A. Juhin, C. Brouder, M.-A. Arrio and Ph. Sainctavit. We acknowledge the support of the Laue lab at the Paul Scherrer Institut (PSI). We thank Thomas Schmidt for his valuable assistance with the experiment. The synchrotron experiments have been performed at the ADRESS beamline of the Swiss Light Source (SLS) at the PSI. The work at PSI is supported by the Swiss National Science Foundation through the NCCR MARVEL and the Sinergia network Mott Physics Beyond the Heisenberg Model (MPBH). M. Sikora acknowledges support from National Science Center of Poland (2014/14/E/ST3/00026). The research in Dresden is supported  by the Deutsche Forschungsgemeinschaft through Grant No. 320571839. This work was financed by the ERC advanced Grant XRAYonACTIVE No. 340279. 

\bibliographystyle{apsrev}

\bibliography{./Citation/CitationLibrary1}

\begin{thebibliography}{49}
\expandafter\ifx\csname natexlab\endcsname\relax\def\natexlab#1{#1}\fi
\expandafter\ifx\csname bibnamefont\endcsname\relax
  \def\bibnamefont#1{#1}\fi
\expandafter\ifx\csname bibfnamefont\endcsname\relax
  \def\bibfnamefont#1{#1}\fi
\expandafter\ifx\csname citenamefont\endcsname\relax
  \def\citenamefont#1{#1}\fi
\expandafter\ifx\csname url\endcsname\relax
  \def\url#1{\texttt{#1}}\fi
\expandafter\ifx\csname urlprefix\endcsname\relax\def\urlprefix{URL }\fi
\providecommand{\bibinfo}[2]{#2}
\providecommand{\eprint}[2][]{\url{#2}}

\bibitem[{\citenamefont{Pontius et~al.}(2011)\citenamefont{Pontius, Kachel,
  Sch{\"u}{\ss}ler-Langeheine, Schlotter, Beye, Sorgenfrei, Chang,
  F{\"o}hlisch, Wurth, Metcalf et~al.}}]{Pontius2011}
\bibinfo{author}{\bibfnamefont{N.}~\bibnamefont{Pontius}},
  \bibinfo{author}{\bibfnamefont{T.}~\bibnamefont{Kachel}},
  \bibinfo{author}{\bibfnamefont{C.}~\bibnamefont{Sch{\"u}{\ss}ler-Langeheine}},
  \bibinfo{author}{\bibfnamefont{W.~F.} \bibnamefont{Schlotter}},
  \bibinfo{author}{\bibfnamefont{M.}~\bibnamefont{Beye}},
  \bibinfo{author}{\bibfnamefont{F.}~\bibnamefont{Sorgenfrei}},
  \bibinfo{author}{\bibfnamefont{C.~F.} \bibnamefont{Chang}},
  \bibinfo{author}{\bibfnamefont{A.}~\bibnamefont{F{\"o}hlisch}},
  \bibinfo{author}{\bibfnamefont{W.}~\bibnamefont{Wurth}},
  \bibinfo{author}{\bibfnamefont{P.}~\bibnamefont{Metcalf}},
  \bibnamefont{et~al.}, \bibinfo{journal}{Appl. Phys. Lett.}
  \textbf{\bibinfo{volume}{98}}, \bibinfo{pages}{182504}
  (\bibinfo{year}{2011}).

\bibitem[{\citenamefont{de~Jong et~al.}(2013)\citenamefont{de~Jong, Kukreja,
  Trabant, Pontius, Chang, Kachel, Beye, Sorgenfrei, Back, Br{\"a}uer
  et~al.}}]{deJong2013}
\bibinfo{author}{\bibfnamefont{S.}~\bibnamefont{de~Jong}},
  \bibinfo{author}{\bibfnamefont{R.}~\bibnamefont{Kukreja}},
  \bibinfo{author}{\bibfnamefont{C.}~\bibnamefont{Trabant}},
  \bibinfo{author}{\bibfnamefont{N.}~\bibnamefont{Pontius}},
  \bibinfo{author}{\bibfnamefont{C.~F.} \bibnamefont{Chang}},
  \bibinfo{author}{\bibfnamefont{T.}~\bibnamefont{Kachel}},
  \bibinfo{author}{\bibfnamefont{M.}~\bibnamefont{Beye}},
  \bibinfo{author}{\bibfnamefont{F.}~\bibnamefont{Sorgenfrei}},
  \bibinfo{author}{\bibfnamefont{C.~H.} \bibnamefont{Back}},
  \bibinfo{author}{\bibfnamefont{B.}~\bibnamefont{Br{\"a}uer}},
  \bibnamefont{et~al.}, \bibinfo{journal}{Nat. Mater}
  \textbf{\bibinfo{volume}{12}}, \bibinfo{pages}{882 EP }
  (\bibinfo{year}{2013}).

\bibitem[{\citenamefont{Verwey}(1939)}]{Verwey1939}
\bibinfo{author}{\bibfnamefont{E.~J.~W.} \bibnamefont{Verwey}},
  \bibinfo{journal}{Nature} \textbf{\bibinfo{volume}{144}}
  (\bibinfo{year}{1939}).

\bibitem[{\citenamefont{Hamilton}(1958)}]{Hamilton1958}
\bibinfo{author}{\bibfnamefont{W.~C.} \bibnamefont{Hamilton}},
  \bibinfo{journal}{Phys. Rev.} \textbf{\bibinfo{volume}{110}},
  \bibinfo{pages}{1050} (\bibinfo{year}{1958}).

\bibitem[{\citenamefont{Yamada et~al.}(1968)\citenamefont{Yamada, Suzuki, and
  Chikazumi}}]{Yamada1968}
\bibinfo{author}{\bibfnamefont{T.}~\bibnamefont{Yamada}},
  \bibinfo{author}{\bibfnamefont{K.}~\bibnamefont{Suzuki}}, \bibnamefont{and}
  \bibinfo{author}{\bibfnamefont{S.}~\bibnamefont{Chikazumi}},
  \bibinfo{journal}{Appl. Phys. Lett.} \textbf{\bibinfo{volume}{13}},
  \bibinfo{pages}{172} (\bibinfo{year}{1968}).

\bibitem[{\citenamefont{Hargrove and Kundig}(1970)}]{Hargrove1970}
\bibinfo{author}{\bibfnamefont{R.}~\bibnamefont{Hargrove}} \bibnamefont{and}
  \bibinfo{author}{\bibfnamefont{W.}~\bibnamefont{Kundig}},
  \bibinfo{journal}{Solid State Commun.} \textbf{\bibinfo{volume}{8}},
  \bibinfo{pages}{303} (\bibinfo{year}{1970}).

\bibitem[{\citenamefont{Gasparov et~al.}(2000)\citenamefont{Gasparov, Tanner,
  Romero, Berger, Margaritondo, and Forr\'o}}]{Gasparov2000}
\bibinfo{author}{\bibfnamefont{L.~V.} \bibnamefont{Gasparov}},
  \bibinfo{author}{\bibfnamefont{D.~B.} \bibnamefont{Tanner}},
  \bibinfo{author}{\bibfnamefont{D.~B.} \bibnamefont{Romero}},
  \bibinfo{author}{\bibfnamefont{H.}~\bibnamefont{Berger}},
  \bibinfo{author}{\bibfnamefont{G.}~\bibnamefont{Margaritondo}},
  \bibnamefont{and} \bibinfo{author}{\bibfnamefont{L.}~\bibnamefont{Forr\'o}},
  \bibinfo{journal}{Phys. Rev. B} \textbf{\bibinfo{volume}{62}},
  \bibinfo{pages}{7939} (\bibinfo{year}{2000}).

\bibitem[{\citenamefont{Nov{\'{a}}k and Brabers}(2000)}]{Novak2000}
\bibinfo{author}{\bibfnamefont{P.}~\bibnamefont{Nov{\'{a}}k}} \bibnamefont{and}
  \bibinfo{author}{\bibfnamefont{V.~A.} \bibnamefont{Brabers}},
  \bibinfo{journal}{Phys. Rev. B} \textbf{\bibinfo{volume}{61}},
  \bibinfo{pages}{1256} (\bibinfo{year}{2000}).

\bibitem[{\citenamefont{Garc{\'{i}}a et~al.}(2001)\citenamefont{Garc{\'{i}}a,
  Sub{\'{i}}as, Proietti, Blasco, Renevier, Hodeau, and Joly}}]{Garcia2001}
\bibinfo{author}{\bibfnamefont{J.}~\bibnamefont{Garc{\'{i}}a}},
  \bibinfo{author}{\bibfnamefont{G.}~\bibnamefont{Sub{\'{i}}as}},
  \bibinfo{author}{\bibfnamefont{M.~G.} \bibnamefont{Proietti}},
  \bibinfo{author}{\bibfnamefont{J.}~\bibnamefont{Blasco}},
  \bibinfo{author}{\bibfnamefont{H.}~\bibnamefont{Renevier}},
  \bibinfo{author}{\bibfnamefont{J.~L.} \bibnamefont{Hodeau}},
  \bibnamefont{and} \bibinfo{author}{\bibfnamefont{Y.}~\bibnamefont{Joly}},
  \bibinfo{journal}{Phys. Rev. B} \textbf{\bibinfo{volume}{63}},
  \bibinfo{pages}{54110} (\bibinfo{year}{2001}).

\bibitem[{\citenamefont{Sub{\'{i}}as et~al.}(2004)\citenamefont{Sub{\'{i}}as,
  Garc{\'{i}}a, Blasco, Proietti, Renevier, and S{\'{a}}nchez}}]{Gracia2004a}
\bibinfo{author}{\bibfnamefont{G.}~\bibnamefont{Sub{\'{i}}as}},
  \bibinfo{author}{\bibfnamefont{J.}~\bibnamefont{Garc{\'{i}}a}},
  \bibinfo{author}{\bibfnamefont{J.}~\bibnamefont{Blasco}},
  \bibinfo{author}{\bibfnamefont{M.~G.} \bibnamefont{Proietti}},
  \bibinfo{author}{\bibfnamefont{H.}~\bibnamefont{Renevier}}, \bibnamefont{and}
  \bibinfo{author}{\bibfnamefont{M.~C.} \bibnamefont{S{\'{a}}nchez}},
  \bibinfo{journal}{Phys. Rev. Lett.} \textbf{\bibinfo{volume}{93}},
  \bibinfo{pages}{15} (\bibinfo{year}{2004}).

\bibitem[{\citenamefont{Senn et~al.}(2011)\citenamefont{Senn, Wright, and
  Attfield}}]{Senn2012b}
\bibinfo{author}{\bibfnamefont{M.~S.} \bibnamefont{Senn}},
  \bibinfo{author}{\bibfnamefont{J.~P.} \bibnamefont{Wright}},
  \bibnamefont{and} \bibinfo{author}{\bibfnamefont{J.~P.}
  \bibnamefont{Attfield}}, \bibinfo{journal}{Nature}
  \textbf{\bibinfo{volume}{481}}, \bibinfo{pages}{173} (\bibinfo{year}{2011}).

\bibitem[{\citenamefont{Senn et~al.}(2013)\citenamefont{Senn, Wright, and
  Attfield}}]{Senn2013}
\bibinfo{author}{\bibfnamefont{M.~S.} \bibnamefont{Senn}},
  \bibinfo{author}{\bibfnamefont{J.~P.} \bibnamefont{Wright}},
  \bibnamefont{and} \bibinfo{author}{\bibfnamefont{J.~P.}
  \bibnamefont{Attfield}}, \bibinfo{journal}{J. Korean Phys. Soc.}
  \textbf{\bibinfo{volume}{62}}, \bibinfo{pages}{1372} (\bibinfo{year}{2013}).

\bibitem[{\citenamefont{Bosak et~al.}(2014)\citenamefont{Bosak, Chernyshov,
  Hoesch, Piekarz, Tacon, Krisch, Kozlowski, Oles, and Parlinski}}]{Bosak2014}
\bibinfo{author}{\bibfnamefont{A.}~\bibnamefont{Bosak}},
  \bibinfo{author}{\bibfnamefont{D.}~\bibnamefont{Chernyshov}},
  \bibinfo{author}{\bibfnamefont{M.}~\bibnamefont{Hoesch}},
  \bibinfo{author}{\bibfnamefont{P.}~\bibnamefont{Piekarz}},
  \bibinfo{author}{\bibfnamefont{M.~L.} \bibnamefont{Tacon}},
  \bibinfo{author}{\bibfnamefont{M.}~\bibnamefont{Krisch}},
  \bibinfo{author}{\bibfnamefont{A.}~\bibnamefont{Kozlowski}},
  \bibinfo{author}{\bibfnamefont{A.~M.} \bibnamefont{Oles}}, \bibnamefont{and}
  \bibinfo{author}{\bibfnamefont{K.}~\bibnamefont{Parlinski}},
  \bibinfo{journal}{Phys. Rev. X} \textbf{\bibinfo{volume}{4}},
  \bibinfo{pages}{1} (\bibinfo{year}{2014}).

\bibitem[{\citenamefont{Taguchi et~al.}(2015)\citenamefont{Taguchi, Chainani,
  Ueda, Matsunami, Ishida, Eguchi, Tsuda, Takata, Yabashi, Tamasaku
  et~al.}}]{Shin2015}
\bibinfo{author}{\bibfnamefont{M.}~\bibnamefont{Taguchi}},
  \bibinfo{author}{\bibfnamefont{A.}~\bibnamefont{Chainani}},
  \bibinfo{author}{\bibfnamefont{S.}~\bibnamefont{Ueda}},
  \bibinfo{author}{\bibfnamefont{M.}~\bibnamefont{Matsunami}},
  \bibinfo{author}{\bibfnamefont{Y.}~\bibnamefont{Ishida}},
  \bibinfo{author}{\bibfnamefont{R.}~\bibnamefont{Eguchi}},
  \bibinfo{author}{\bibfnamefont{S.}~\bibnamefont{Tsuda}},
  \bibinfo{author}{\bibfnamefont{Y.}~\bibnamefont{Takata}},
  \bibinfo{author}{\bibfnamefont{M.}~\bibnamefont{Yabashi}},
  \bibinfo{author}{\bibfnamefont{K.}~\bibnamefont{Tamasaku}},
  \bibnamefont{et~al.}, \bibinfo{journal}{Phys.Rev. Lett.}
  \textbf{\bibinfo{volume}{115}}, \bibinfo{pages}{1} (\bibinfo{year}{2015}).

\bibitem[{\citenamefont{Huang et~al.}(2017)\citenamefont{Huang, Chen, Wang, {De
  Groot}, Wu, Okamoto, Chainani, Zhou, Jeng, Guo et~al.}}]{Huang2017}
\bibinfo{author}{\bibfnamefont{H.~Y.} \bibnamefont{Huang}},
  \bibinfo{author}{\bibfnamefont{Z.~Y.} \bibnamefont{Chen}},
  \bibinfo{author}{\bibfnamefont{R.~P.} \bibnamefont{Wang}},
  \bibinfo{author}{\bibfnamefont{F.~M.} \bibnamefont{{De Groot}}},
  \bibinfo{author}{\bibfnamefont{W.~B.} \bibnamefont{Wu}},
  \bibinfo{author}{\bibfnamefont{J.}~\bibnamefont{Okamoto}},
  \bibinfo{author}{\bibfnamefont{A.}~\bibnamefont{Chainani}},
  \bibinfo{author}{\bibfnamefont{J.~S.} \bibnamefont{Zhou}},
  \bibinfo{author}{\bibfnamefont{H.~T.} \bibnamefont{Jeng}},
  \bibinfo{author}{\bibfnamefont{G.~Y.} \bibnamefont{Guo}},
  \bibnamefont{et~al.}, \bibinfo{journal}{Nat. Commun.}
  \textbf{\bibinfo{volume}{9}}, \bibinfo{pages}{0} (\bibinfo{year}{2017}).

\bibitem[{\citenamefont{Walz}(2002)}]{Walz2002}
\bibinfo{author}{\bibfnamefont{F.}~\bibnamefont{Walz}}, \bibinfo{journal}{J.
  Phys. Condens. Matter} \textbf{\bibinfo{volume}{14}} (\bibinfo{year}{2002}).

\bibitem[{\citenamefont{Garc{\'{i}}a and Sub{\'{i}}as}(2004)}]{Garcia2004}
\bibinfo{author}{\bibfnamefont{J.}~\bibnamefont{Garc{\'{i}}a}}
  \bibnamefont{and}
  \bibinfo{author}{\bibfnamefont{G.}~\bibnamefont{Sub{\'{i}}as}},
  \bibinfo{journal}{J. Phys. Condens. Matter} \textbf{\bibinfo{volume}{16}},
  \bibinfo{pages}{R145} (\bibinfo{year}{2004}).

\bibitem[{\citenamefont{Ament et~al.}(2011)\citenamefont{Ament, van Veenendaal,
  Devereaux, Hill, and van~den Brink}}]{Ament2011}
\bibinfo{author}{\bibfnamefont{L.~J.~P.} \bibnamefont{Ament}},
  \bibinfo{author}{\bibfnamefont{M.}~\bibnamefont{van Veenendaal}},
  \bibinfo{author}{\bibfnamefont{T.~P.} \bibnamefont{Devereaux}},
  \bibinfo{author}{\bibfnamefont{J.~P.} \bibnamefont{Hill}}, \bibnamefont{and}
  \bibinfo{author}{\bibfnamefont{J.}~\bibnamefont{van~den Brink}},
  \bibinfo{journal}{Rev. Mod. Phys.} \textbf{\bibinfo{volume}{83}},
  \bibinfo{pages}{705} (\bibinfo{year}{2011}).

\bibitem[{\citenamefont{Bisogni et~al.}(2016)\citenamefont{Bisogni, Catalano,
  Green, Gibert, Scherwitzl, Huang, Strocov, Zubko, Balandeh, Triscone
  et~al.}}]{Bisogni2016a}
\bibinfo{author}{\bibfnamefont{V.}~\bibnamefont{Bisogni}},
  \bibinfo{author}{\bibfnamefont{S.}~\bibnamefont{Catalano}},
  \bibinfo{author}{\bibfnamefont{R.~J.} \bibnamefont{Green}},
  \bibinfo{author}{\bibfnamefont{M.}~\bibnamefont{Gibert}},
  \bibinfo{author}{\bibfnamefont{R.}~\bibnamefont{Scherwitzl}},
  \bibinfo{author}{\bibfnamefont{Y.}~\bibnamefont{Huang}},
  \bibinfo{author}{\bibfnamefont{V.~N.} \bibnamefont{Strocov}},
  \bibinfo{author}{\bibfnamefont{P.}~\bibnamefont{Zubko}},
  \bibinfo{author}{\bibfnamefont{S.}~\bibnamefont{Balandeh}},
  \bibinfo{author}{\bibfnamefont{J.~M.} \bibnamefont{Triscone}},
  \bibnamefont{et~al.}, \bibinfo{journal}{Nat. Commun.}
  \textbf{\bibinfo{volume}{7}}, \bibinfo{pages}{1} (\bibinfo{year}{2016}).

\bibitem[{\citenamefont{Lu et~al.}(2018)\citenamefont{Lu, Betto, F{\"u}rsich,
  Suzuki, Kim, Cristiani, Logvenov, Brookes, Benckiser, Haverkort
  et~al.}}]{Lu2018}
\bibinfo{author}{\bibfnamefont{Y.}~\bibnamefont{Lu}},
  \bibinfo{author}{\bibfnamefont{D.}~\bibnamefont{Betto}},
  \bibinfo{author}{\bibfnamefont{K.}~\bibnamefont{F{\"u}rsich}},
  \bibinfo{author}{\bibfnamefont{H.}~\bibnamefont{Suzuki}},
  \bibinfo{author}{\bibfnamefont{H.~H.} \bibnamefont{Kim}},
  \bibinfo{author}{\bibfnamefont{G.}~\bibnamefont{Cristiani}},
  \bibinfo{author}{\bibfnamefont{G.}~\bibnamefont{Logvenov}},
  \bibinfo{author}{\bibfnamefont{N.~B.} \bibnamefont{Brookes}},
  \bibinfo{author}{\bibfnamefont{E.}~\bibnamefont{Benckiser}},
  \bibinfo{author}{\bibfnamefont{M.~W.} \bibnamefont{Haverkort}},
  \bibnamefont{et~al.}, \bibinfo{journal}{Phys. Rev. X}
  (\bibinfo{year}{2018}).

\bibitem[{\citenamefont{Miyawaki et~al.}(2017)\citenamefont{Miyawaki, Suga,
  Fujiwara, Urasaki, Ikeno, Niwa, Kiuchi, and Harada}}]{Miyawaki2017}
\bibinfo{author}{\bibfnamefont{J.}~\bibnamefont{Miyawaki}},
  \bibinfo{author}{\bibfnamefont{S.}~\bibnamefont{Suga}},
  \bibinfo{author}{\bibfnamefont{H.}~\bibnamefont{Fujiwara}},
  \bibinfo{author}{\bibfnamefont{M.}~\bibnamefont{Urasaki}},
  \bibinfo{author}{\bibfnamefont{H.}~\bibnamefont{Ikeno}},
  \bibinfo{author}{\bibfnamefont{H.}~\bibnamefont{Niwa}},
  \bibinfo{author}{\bibfnamefont{H.}~\bibnamefont{Kiuchi}}, \bibnamefont{and}
  \bibinfo{author}{\bibfnamefont{Y.}~\bibnamefont{Harada}},
  \bibinfo{journal}{Phys. Rev. B} \textbf{\bibinfo{volume}{96}},
  \bibinfo{pages}{214420} (\bibinfo{year}{2017}).

\bibitem[{sup()}]{supplementary}
\bibinfo{howpublished}{Supplemental Material}.

\bibitem[{\citenamefont{Strocov et~al.}(2010)\citenamefont{Strocov, Schmitt,
  Flechsig, Schmidt, Imhof, Chen, Raabe, Betemps, Zimoch, Krempasky
  et~al.}}]{Strocov2010}
\bibinfo{author}{\bibfnamefont{V.~N.} \bibnamefont{Strocov}},
  \bibinfo{author}{\bibfnamefont{T.}~\bibnamefont{Schmitt}},
  \bibinfo{author}{\bibfnamefont{U.}~\bibnamefont{Flechsig}},
  \bibinfo{author}{\bibfnamefont{T.}~\bibnamefont{Schmidt}},
  \bibinfo{author}{\bibfnamefont{A.}~\bibnamefont{Imhof}},
  \bibinfo{author}{\bibfnamefont{Q.}~\bibnamefont{Chen}},
  \bibinfo{author}{\bibfnamefont{J.}~\bibnamefont{Raabe}},
  \bibinfo{author}{\bibfnamefont{R.}~\bibnamefont{Betemps}},
  \bibinfo{author}{\bibfnamefont{D.}~\bibnamefont{Zimoch}},
  \bibinfo{author}{\bibfnamefont{J.}~\bibnamefont{Krempasky}},
  \bibnamefont{et~al.}, \bibinfo{journal}{‎J. Synchrotron Radiat.}
  \textbf{\bibinfo{volume}{17}}, \bibinfo{pages}{631} (\bibinfo{year}{2010}).

\bibitem[{\citenamefont{Sasaki et~al.}(1993)\citenamefont{Sasaki, Kakuno,
  Takada, Shimada, ichi Yanagida, and Miyahara}}]{Sasaki1993}
\bibinfo{author}{\bibfnamefont{S.}~\bibnamefont{Sasaki}},
  \bibinfo{author}{\bibfnamefont{K.}~\bibnamefont{Kakuno}},
  \bibinfo{author}{\bibfnamefont{T.}~\bibnamefont{Takada}},
  \bibinfo{author}{\bibfnamefont{T.}~\bibnamefont{Shimada}},
  \bibinfo{author}{\bibfnamefont{K.}~\bibnamefont{ichi Yanagida}},
  \bibnamefont{and} \bibinfo{author}{\bibfnamefont{Y.}~\bibnamefont{Miyahara}},
  \bibinfo{journal}{Nucl. Instrum. Methods Phys. Res.}
  \textbf{\bibinfo{volume}{331}}, \bibinfo{pages}{763 } (\bibinfo{year}{1993}).

\bibitem[{\citenamefont{Haverkort et~al.}(2012)\citenamefont{Haverkort,
  Zwierzycki, and Andersen}}]{Haverkort2012}
\bibinfo{author}{\bibfnamefont{M.~W.} \bibnamefont{Haverkort}},
  \bibinfo{author}{\bibfnamefont{M.}~\bibnamefont{Zwierzycki}},
  \bibnamefont{and} \bibinfo{author}{\bibfnamefont{O.~K.}
  \bibnamefont{Andersen}}, \bibinfo{journal}{Phys. Rev. B}
  \textbf{\bibinfo{volume}{85}}, \bibinfo{pages}{1} (\bibinfo{year}{2012}).

\bibitem[{\citenamefont{Lua et~al.}(2014)\citenamefont{Lua, Hoeppner,
  Gunnarsson, and Haverkort}}]{Lu2014}
\bibinfo{author}{\bibfnamefont{Y.}~\bibnamefont{Lua}},
  \bibinfo{author}{\bibfnamefont{M.}~\bibnamefont{Hoeppner}},
  \bibinfo{author}{\bibfnamefont{O.}~\bibnamefont{Gunnarsson}},
  \bibnamefont{and} \bibinfo{author}{\bibfnamefont{M.~W.}
  \bibnamefont{Haverkort}}, \bibinfo{journal}{Phys. Rev. B}
  \textbf{\bibinfo{volume}{90}}, \bibinfo{pages}{085102}
  (\bibinfo{year}{2014}).

\bibitem[{\citenamefont{Haverkort et~al.}(2014)\citenamefont{Haverkort,
  Sangiovanni, Hansmann, Toschi, Lu, and Macke}}]{Haverkort2014}
\bibinfo{author}{\bibfnamefont{M.~W.} \bibnamefont{Haverkort}},
  \bibinfo{author}{\bibfnamefont{G.}~\bibnamefont{Sangiovanni}},
  \bibinfo{author}{\bibfnamefont{P.}~\bibnamefont{Hansmann}},
  \bibinfo{author}{\bibfnamefont{A.}~\bibnamefont{Toschi}},
  \bibinfo{author}{\bibfnamefont{Y.}~\bibnamefont{Lu}}, \bibnamefont{and}
  \bibinfo{author}{\bibfnamefont{S.}~\bibnamefont{Macke}},
  \bibinfo{journal}{EPL} \textbf{\bibinfo{volume}{108}}, \bibinfo{pages}{57004}
  (\bibinfo{year}{2014}).

\bibitem[{\citenamefont{Grave et~al.}(1993)\citenamefont{Grave, Persoons,
  Vandenberghe, and Bakker}}]{Grave1993}
\bibinfo{author}{\bibfnamefont{E.~D.} \bibnamefont{Grave}},
  \bibinfo{author}{\bibfnamefont{R.~M.} \bibnamefont{Persoons}},
  \bibinfo{author}{\bibfnamefont{R.~E.} \bibnamefont{Vandenberghe}},
  \bibnamefont{and} \bibinfo{author}{\bibfnamefont{P.~M. A.~D.}
  \bibnamefont{Bakker}}, \bibinfo{journal}{Phys. Rev. B}
  \textbf{\bibinfo{volume}{47}} (\bibinfo{year}{1993}).

\bibitem[{\citenamefont{Kuiper et~al.}(1997)\citenamefont{Kuiper, Searle, Duda,
  and van~der Zaag}}]{Kuiper1997}
\bibinfo{author}{\bibfnamefont{P.}~\bibnamefont{Kuiper}},
  \bibinfo{author}{\bibfnamefont{G.}~\bibnamefont{Searle}},
  \bibinfo{author}{\bibfnamefont{L.}~\bibnamefont{Duda}}, \bibnamefont{and}
  \bibinfo{author}{\bibnamefont{van~der Zaag}}, \bibinfo{journal}{J. Electron
  Spectros. Relat. Phenomena.} \textbf{\bibinfo{volume}{86}}
  (\bibinfo{year}{1997}).

\bibitem[{\citenamefont{Pattrick et~al.}(2002)\citenamefont{Pattrick, {Van Der
  Laan}, Henderson, Kuiper, Dudzik, and Vaughan}}]{Pattrick2002}
\bibinfo{author}{\bibfnamefont{R.~a.~D.} \bibnamefont{Pattrick}},
  \bibinfo{author}{\bibfnamefont{G.}~\bibnamefont{{Van Der Laan}}},
  \bibinfo{author}{\bibfnamefont{C.~M.~B.} \bibnamefont{Henderson}},
  \bibinfo{author}{\bibfnamefont{P.}~\bibnamefont{Kuiper}},
  \bibinfo{author}{\bibfnamefont{E.}~\bibnamefont{Dudzik}}, \bibnamefont{and}
  \bibinfo{author}{\bibfnamefont{D.~J.} \bibnamefont{Vaughan}},
  \bibinfo{journal}{Eur. J. Mineral.} \textbf{\bibinfo{volume}{14}},
  \bibinfo{pages}{1095} (\bibinfo{year}{2002}).

\bibitem[{\citenamefont{van~der Laan et~al.}(2006)\citenamefont{van~der Laan,
  Arenholz, van~der Laan, Chopdekar, Suzuki, and Rajesh}}]{Arenholz2006a}
\bibinfo{author}{\bibfnamefont{G.}~\bibnamefont{van~der Laan}},
  \bibinfo{author}{\bibfnamefont{E.}~\bibnamefont{Arenholz}},
  \bibinfo{author}{\bibfnamefont{G.}~\bibnamefont{van~der Laan}},
  \bibinfo{author}{\bibfnamefont{R.~V.} \bibnamefont{Chopdekar}},
  \bibinfo{author}{\bibfnamefont{Y.}~\bibnamefont{Suzuki}}, \bibnamefont{and}
  \bibinfo{author}{\bibfnamefont{V.}~\bibnamefont{Rajesh}},
  \bibinfo{journal}{Phys. Rev. B} \textbf{\bibinfo{volume}{74}},
  \bibinfo{pages}{1} (\bibinfo{year}{2006}).

\bibitem[{\citenamefont{Carvallo et~al.}(2008)\citenamefont{Carvallo,
  Sainctavit, Arrio, Menguy, Wang, Ona-Nguema, and
  Brice-Profeta}}]{Carvallo2008}
\bibinfo{author}{\bibfnamefont{C.}~\bibnamefont{Carvallo}},
  \bibinfo{author}{\bibfnamefont{P.}~\bibnamefont{Sainctavit}},
  \bibinfo{author}{\bibfnamefont{M.~A.} \bibnamefont{Arrio}},
  \bibinfo{author}{\bibfnamefont{N.}~\bibnamefont{Menguy}},
  \bibinfo{author}{\bibfnamefont{Y.}~\bibnamefont{Wang}},
  \bibinfo{author}{\bibfnamefont{G.}~\bibnamefont{Ona-Nguema}},
  \bibnamefont{and}
  \bibinfo{author}{\bibfnamefont{S.}~\bibnamefont{Brice-Profeta}},
  \bibinfo{journal}{Am. Mineral.} \textbf{\bibinfo{volume}{93}},
  \bibinfo{pages}{880} (\bibinfo{year}{2008}).

\bibitem[{\citenamefont{Liu et~al.}(2017)\citenamefont{Liu, Piamonteze,
  Delgado-jaime, Wang, Dreiser, Chopdekar, Nolting, and {De Groot}}}]{Liu2017}
\bibinfo{author}{\bibfnamefont{B.}~\bibnamefont{Liu}},
  \bibinfo{author}{\bibfnamefont{C.}~\bibnamefont{Piamonteze}},
  \bibinfo{author}{\bibfnamefont{M.~U.} \bibnamefont{Delgado-jaime}},
  \bibinfo{author}{\bibfnamefont{R.~P.} \bibnamefont{Wang}},
  \bibinfo{author}{\bibfnamefont{J.}~\bibnamefont{Dreiser}},
  \bibinfo{author}{\bibfnamefont{R.}~\bibnamefont{Chopdekar}},
  \bibinfo{author}{\bibfnamefont{F.}~\bibnamefont{Nolting}}, \bibnamefont{and}
  \bibinfo{author}{\bibfnamefont{F.~M.} \bibnamefont{{De Groot}}},
  \bibinfo{journal}{Phys. Rev. B} \textbf{\bibinfo{volume}{96}},
  \bibinfo{pages}{1} (\bibinfo{year}{2017}).

\bibitem[{\citenamefont{Goering et~al.}(2005)\citenamefont{Goering, Gold,
  Lafkioti, and Sch{\"{u}}tz}}]{Goering2005}
\bibinfo{author}{\bibfnamefont{E.}~\bibnamefont{Goering}},
  \bibinfo{author}{\bibfnamefont{S.}~\bibnamefont{Gold}},
  \bibinfo{author}{\bibfnamefont{M.}~\bibnamefont{Lafkioti}}, \bibnamefont{and}
  \bibinfo{author}{\bibfnamefont{G.}~\bibnamefont{Sch{\"{u}}tz}},
  \bibinfo{journal}{Europhys. Lett.} \textbf{\bibinfo{volume}{73}},
  \bibinfo{pages}{7} (\bibinfo{year}{2005}).

\bibitem[{\citenamefont{Goering et~al.}(2007)\citenamefont{Goering, Lafkioti,
  Gold, and Sch{\"{u}}tz}}]{Goering2007}
\bibinfo{author}{\bibfnamefont{E.}~\bibnamefont{Goering}},
  \bibinfo{author}{\bibfnamefont{M.}~\bibnamefont{Lafkioti}},
  \bibinfo{author}{\bibfnamefont{S.}~\bibnamefont{Gold}}, \bibnamefont{and}
  \bibinfo{author}{\bibfnamefont{G.}~\bibnamefont{Sch{\"{u}}tz}},
  \bibinfo{journal}{J. Magn. Magn. Mater.} \textbf{\bibinfo{volume}{310}},
  \bibinfo{pages}{249} (\bibinfo{year}{2007}).

\bibitem[{\citenamefont{Antonov et~al.}(2003)\citenamefont{Antonov, Harmon, and
  Yaresko}}]{Antonov2003}
\bibinfo{author}{\bibfnamefont{V.~N.} \bibnamefont{Antonov}},
  \bibinfo{author}{\bibfnamefont{B.~N.} \bibnamefont{Harmon}},
  \bibnamefont{and} \bibinfo{author}{\bibfnamefont{A.~N.}
  \bibnamefont{Yaresko}}, \bibinfo{journal}{Phys. Rev. B}
  \textbf{\bibinfo{volume}{67}}, \bibinfo{pages}{024417}
  (\bibinfo{year}{2003}).

\bibitem[{\citenamefont{Juhin et~al.}(2014)\citenamefont{Juhin, Brouder, and
  {De Groot}}}]{Juhin2014}
\bibinfo{author}{\bibfnamefont{A.}~\bibnamefont{Juhin}},
  \bibinfo{author}{\bibfnamefont{C.}~\bibnamefont{Brouder}}, \bibnamefont{and}
  \bibinfo{author}{\bibfnamefont{F.~M.} \bibnamefont{{De Groot}}},
  \bibinfo{journal}{Cent. Eur. J. Phys.} \textbf{\bibinfo{volume}{12}},
  \bibinfo{pages}{323} (\bibinfo{year}{2014}).

\bibitem[{\citenamefont{Haverkort}(2010)}]{Haverkort2010}
\bibinfo{author}{\bibfnamefont{M.~W.} \bibnamefont{Haverkort}},
  \bibinfo{journal}{Phys. Rev. Lett.} \textbf{\bibinfo{volume}{105}},
  \bibinfo{pages}{167404} (\bibinfo{year}{2010}).

\bibitem[{\citenamefont{M.F. et~al.}(1998)\citenamefont{M.F., Kuiper, and
  Sawatzky}}]{DeGroot1998}
\bibinfo{author}{\bibfnamefont{F.}~\bibnamefont{M.F.}},
  \bibinfo{author}{\bibfnamefont{P.}~\bibnamefont{Kuiper}}, \bibnamefont{and}
  \bibinfo{author}{\bibfnamefont{G.~A.} \bibnamefont{Sawatzky}},
  \bibinfo{journal}{Phys. Rev. B} \textbf{\bibinfo{volume}{57}},
  \bibinfo{pages}{14584} (\bibinfo{year}{1998}).

\bibitem[{\citenamefont{Betto et~al.}(2017)\citenamefont{Betto, Peng, Porter,
  Berti, Calloni, Ghiringhelli, and Brookes}}]{Betto2017}
\bibinfo{author}{\bibfnamefont{D.}~\bibnamefont{Betto}},
  \bibinfo{author}{\bibfnamefont{Y.~Y.} \bibnamefont{Peng}},
  \bibinfo{author}{\bibfnamefont{S.~B.} \bibnamefont{Porter}},
  \bibinfo{author}{\bibfnamefont{G.}~\bibnamefont{Berti}},
  \bibinfo{author}{\bibfnamefont{A.}~\bibnamefont{Calloni}},
  \bibinfo{author}{\bibfnamefont{G.}~\bibnamefont{Ghiringhelli}},
  \bibnamefont{and} \bibinfo{author}{\bibfnamefont{N.~B.}
  \bibnamefont{Brookes}}, \bibinfo{journal}{Phys. Rev. B}
  \textbf{\bibinfo{volume}{96}} (\bibinfo{year}{2017}).

\bibitem[{\citenamefont{Goering}(2011)}]{Goering2011}
\bibinfo{author}{\bibfnamefont{E.}~\bibnamefont{Goering}},
  \bibinfo{journal}{Phys. Status Solidi B} \textbf{\bibinfo{volume}{248}},
  \bibinfo{pages}{2345} (\bibinfo{year}{2011}).

\bibitem[{\citenamefont{McQueeney et~al.}(2007)\citenamefont{McQueeney,
  Yethiraj, Chang, Montfrooij, Perring, Honig, and Metcalf}}]{McQueeney2007}
\bibinfo{author}{\bibfnamefont{R.~J.} \bibnamefont{McQueeney}},
  \bibinfo{author}{\bibfnamefont{M.}~\bibnamefont{Yethiraj}},
  \bibinfo{author}{\bibfnamefont{S.}~\bibnamefont{Chang}},
  \bibinfo{author}{\bibfnamefont{W.}~\bibnamefont{Montfrooij}},
  \bibinfo{author}{\bibfnamefont{T.~G.} \bibnamefont{Perring}},
  \bibinfo{author}{\bibfnamefont{J.~M.} \bibnamefont{Honig}}, \bibnamefont{and}
  \bibinfo{author}{\bibfnamefont{P.}~\bibnamefont{Metcalf}},
  \bibinfo{journal}{Phys. Rev. Lett.} \textbf{\bibinfo{volume}{99}},
  \bibinfo{pages}{246401} (\bibinfo{year}{2007}).

\bibitem[{\citenamefont{Ghiringhelli et~al.}(2009)\citenamefont{Ghiringhelli,
  Piazzalunga, Dallera, Schmitt, Strocov, Schlappa, Patthey, Wang, Berger, and
  Grioni}}]{Ghiringhelli2009}
\bibinfo{author}{\bibfnamefont{G.}~\bibnamefont{Ghiringhelli}},
  \bibinfo{author}{\bibfnamefont{A.}~\bibnamefont{Piazzalunga}},
  \bibinfo{author}{\bibfnamefont{C.}~\bibnamefont{Dallera}},
  \bibinfo{author}{\bibfnamefont{T.}~\bibnamefont{Schmitt}},
  \bibinfo{author}{\bibfnamefont{V.~N.} \bibnamefont{Strocov}},
  \bibinfo{author}{\bibfnamefont{J.}~\bibnamefont{Schlappa}},
  \bibinfo{author}{\bibfnamefont{L.}~\bibnamefont{Patthey}},
  \bibinfo{author}{\bibfnamefont{X.}~\bibnamefont{Wang}},
  \bibinfo{author}{\bibfnamefont{H.}~\bibnamefont{Berger}}, \bibnamefont{and}
  \bibinfo{author}{\bibfnamefont{M.}~\bibnamefont{Grioni}},
  \bibinfo{journal}{Phys. Rev. Lett.} \textbf{\bibinfo{volume}{102}},
  \bibinfo{pages}{2} (\bibinfo{year}{2009}).

\bibitem[{\citenamefont{Garc{\'{i}}a et~al.}(2000)\citenamefont{Garc{\'{i}}a,
  Sub{\'{i}}as, Proietti, Renevier, Joly, Hodeau, Blasco, S{\'{a}}nchez, and
  B{\'{e}}rar}}]{Garcia2000}
\bibinfo{author}{\bibfnamefont{J.}~\bibnamefont{Garc{\'{i}}a}},
  \bibinfo{author}{\bibfnamefont{G.}~\bibnamefont{Sub{\'{i}}as}},
  \bibinfo{author}{\bibfnamefont{M.~G.} \bibnamefont{Proietti}},
  \bibinfo{author}{\bibfnamefont{H.}~\bibnamefont{Renevier}},
  \bibinfo{author}{\bibfnamefont{Y.}~\bibnamefont{Joly}},
  \bibinfo{author}{\bibfnamefont{J.~L.} \bibnamefont{Hodeau}},
  \bibinfo{author}{\bibfnamefont{J.}~\bibnamefont{Blasco}},
  \bibinfo{author}{\bibfnamefont{M.~C.} \bibnamefont{S{\'{a}}nchez}},
  \bibnamefont{and} \bibinfo{author}{\bibfnamefont{J.~F.}
  \bibnamefont{B{\'{e}}rar}}, \bibinfo{journal}{Phys. Rev. Lett.}
  \textbf{\bibinfo{volume}{85}}, \bibinfo{pages}{578} (\bibinfo{year}{2000}).

\bibitem[{\citenamefont{Haupricht et~al.}(2010)\citenamefont{Haupricht,
  Sutarto, Haverkort, Ott, Tanaka, Hsieh, Lin, Chen, Hu, and
  Tjeng}}]{Haupricht2010}
\bibinfo{author}{\bibfnamefont{T.}~\bibnamefont{Haupricht}},
  \bibinfo{author}{\bibfnamefont{R.}~\bibnamefont{Sutarto}},
  \bibinfo{author}{\bibfnamefont{M.~W.} \bibnamefont{Haverkort}},
  \bibinfo{author}{\bibfnamefont{H.}~\bibnamefont{Ott}},
  \bibinfo{author}{\bibfnamefont{A.}~\bibnamefont{Tanaka}},
  \bibinfo{author}{\bibfnamefont{H.~H.} \bibnamefont{Hsieh}},
  \bibinfo{author}{\bibfnamefont{H.~J.} \bibnamefont{Lin}},
  \bibinfo{author}{\bibfnamefont{C.~T.} \bibnamefont{Chen}},
  \bibinfo{author}{\bibfnamefont{Z.}~\bibnamefont{Hu}}, \bibnamefont{and}
  \bibinfo{author}{\bibfnamefont{L.~H.} \bibnamefont{Tjeng}},
  \bibinfo{journal}{Phys. Rev. B} \textbf{\bibinfo{volume}{82}},
  \bibinfo{pages}{035120} (\bibinfo{year}{2010}), ISSN
  \bibinfo{issn}{1098-0121}.

\bibitem[{\citenamefont{Yamada et~al.}(1979)\citenamefont{Yamada, Mori, and
  Noda}}]{Yamada1979}
\bibinfo{author}{\bibfnamefont{Y.}~\bibnamefont{Yamada}},
  \bibinfo{author}{\bibfnamefont{M.}~\bibnamefont{Mori}}, \bibnamefont{and}
  \bibinfo{author}{\bibfnamefont{Y.}~\bibnamefont{Noda}},
  \bibinfo{journal}{Solid State Commun.} \textbf{\bibinfo{volume}{32}},
  \bibinfo{pages}{827} (\bibinfo{year}{1979}).

\bibitem[{\citenamefont{Uzu and Tanaka}(2006)}]{Uzu2006}
\bibinfo{author}{\bibfnamefont{H.}~\bibnamefont{Uzu}} \bibnamefont{and}
  \bibinfo{author}{\bibfnamefont{A.}~\bibnamefont{Tanaka}},
  \bibinfo{journal}{Physica B} \textbf{\bibinfo{volume}{378-380}},
  \bibinfo{pages}{571} (\bibinfo{year}{2006}).

\bibitem[{\citenamefont{Uzu and Tanaka}(2008)}]{Uzu2008}
\bibinfo{author}{\bibfnamefont{H.}~\bibnamefont{Uzu}} \bibnamefont{and}
  \bibinfo{author}{\bibfnamefont{A.}~\bibnamefont{Tanaka}},
  \bibinfo{journal}{J. Phys. Soc. Jpn.} \textbf{\bibinfo{volume}{77}},
  \bibinfo{pages}{1} (\bibinfo{year}{2008}).

\bibitem[{\citenamefont{Tanaka et~al.}(2012)\citenamefont{Tanaka, Chang,
  Buchholz, Trabant, Schierle, Schlappa, Schmitz, Ott, Metcalf, Tjeng
  et~al.}}]{Tanaka2012}
\bibinfo{author}{\bibfnamefont{A.}~\bibnamefont{Tanaka}},
  \bibinfo{author}{\bibfnamefont{C.~F.} \bibnamefont{Chang}},
  \bibinfo{author}{\bibfnamefont{M.}~\bibnamefont{Buchholz}},
  \bibinfo{author}{\bibfnamefont{C.}~\bibnamefont{Trabant}},
  \bibinfo{author}{\bibfnamefont{E.}~\bibnamefont{Schierle}},
  \bibinfo{author}{\bibfnamefont{J.}~\bibnamefont{Schlappa}},
  \bibinfo{author}{\bibfnamefont{D.}~\bibnamefont{Schmitz}},
  \bibinfo{author}{\bibfnamefont{H.}~\bibnamefont{Ott}},
  \bibinfo{author}{\bibfnamefont{P.}~\bibnamefont{Metcalf}},
  \bibinfo{author}{\bibfnamefont{L.~H.} \bibnamefont{Tjeng}},
  \bibnamefont{et~al.}, \bibinfo{journal}{Phys. Rev. Lett.}
  \textbf{\bibinfo{volume}{108}}, \bibinfo{pages}{1} (\bibinfo{year}{2012}).

\end{thebibliography}
\end{document}